\documentclass[conference]{IEEEtran}
\IEEEoverridecommandlockouts
\usepackage{amsthm}
\usepackage{cite}
\usepackage{fancyhdr}
\usepackage{amsmath,amssymb,amsfonts}
\usepackage{algorithmic}
\usepackage{graphicx}
\usepackage{textcomp}
\usepackage{xcolor}
\usepackage{tabularx}
\usepackage{subcaption}
\usepackage[norelsize]{algorithm2e}
\newtheorem{mydef}{Definition}
\def\BibTeX{{\rm B\kern-.05em{\sc i\kern-.025em b}\kern-.08em
    T\kern-.1667em\lower.7ex\hbox{E}\kern-.125emX}}
\begin{document}

\title{Neural Cryptanalysis: Metrics, Methodology, and Applications in CPS Ciphers}

\author{\IEEEauthorblockN{ Ya Xiao, Qingying Hao and Danfeng (Daphne) Yao}
\IEEEauthorblockA{\textit{Computer Science, Virginia Tech} \\
Blacksburg, VA \\
\{yax99, qhao2018, danfeng\}@vt.edu}
}

\maketitle
\thispagestyle{fancy}

\sloppy

\begin{abstract}
Many real-world cyber-physical systems (CPS) use proprietary cipher algorithms. In this work, we describe an easy-to-use black-box security evaluation approach to measure the strength of proprietary ciphers without having to know the algorithms. We quantify the strength of a cipher by measuring how difficult it is for a neural network to mimic the cipher algorithm. We define new metrics (e.g., cipher match rate, training data complexity and training time complexity) that are computed from neural networks to quantitatively represent the cipher strength. This measurement approach allows us to directly compare the security of ciphers. Our experimental demonstration utilizes fully connected neural networks with multiple parallel binary classifiers at the output layer. The results show that when compared with round-reduced DES, the security strength of Hitag2 (a popular stream cipher used in the keyless entry of modern cars) is weaker than 3-round DES.
\end{abstract}


\section{Introduction}





\begin{figure}[tbh]
    \centering
    \includegraphics[width=0.28\textwidth]{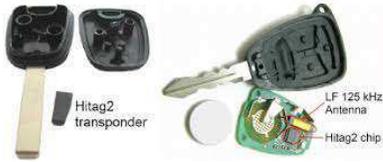}
    \caption{Car keys with Hitag2 transponders shown in \cite{verdult2012gone}}
    \label{fig:hitag2-intro}
\end{figure}

For many real-world cyber-physical systems (e.g. keyless entry of modern vehicles in Figure~\ref{fig:hitag2-intro}), it is common to use lightweight symmetric ciphers. However, unlike the asymmetric ciphers whose security can be reduced to the hardness of mathematical problems, the security of symmetric ciphers is required to be evaluated empirically. Cryptanalysis (e.g.,~\cite{biham1991differential,matsui1993linear,phan2004impossible,biryukov2009related}) refers to the line of work that systematically measures the strength of ciphers.  Cryptanalysis of symmetric ciphers works by launching various attacks and assessing whether or not a cryptographic primitive is resistant to these attacks. 

In traditional symmetric cryptanalysis, an attack is regarded as successful if the key can be recovered with a complexity less than the brute-force key search. There are several existing attack approaches (e.g., linear cryptoanalysis~\cite{matsui1993linear} and differential cryptoanalysis~\cite{biham1992differential}).  A cryptanalyst has to launch them one by one to evaluate a cipher. The security level of the cipher is determined by the best-effort attack. For conventional cryptanalysis, human intervention and manual effort play a central role -- an intrinsic limitation.  Because a feasible attack is enabled by certain unbalanced mathematical relationships (e.g. linear approximation bias, high-probability differential path). These unbalanced mathematical relationships need to be manually identified. Due to the huge search space, one cannot automatically exhaust all possible paths to identify high probabilistic ones.
Thus, conventional cryptanalysis has limited scalability.

Besides, traditional cryptanalysis methods also require knowledge of the cipher algorithm. 
Ciphers on commercial cyber-physical systems (CPS) are usually proprietary, e.g., Hitag2, Megamos Crypto~\cite{em4170125k}. Traditional cryptanalysis approaches cannot be directly applied. Fully recovering the cipher algorithm through reverse engineering may not always be possible. 

In this paper, we address the problem of black-box and scalable cryptanalysis for symmetric ciphers, specifically how to automatically measure the cipher strengths without the knowledge of cipher algorithms. This problem has not been addressed in the cryptanalysis literature. We define \textbf{neural cryptanalysis} as a cryptanalysis approach that leverages the learning ability of neural networks to measure the strengths of ciphers. 

We train neural networks to mimic cipher algorithms. The stronger the cipher is, the more difficult it is for this cipher to be mimicked.  The training data is a collection of plaintext-ciphertext pairs. The task is to predict ciphertexts on the input of plaintexts. This task is equivalent to its opposite version, predicting plaintexts from ciphertexts due to the symmetry of encryption and decryption. The mimicking success breaks the cipher by uncovering the mapping between plaintexts and ciphertexts.  We represent the mimic difficulty by prediction accuracy and corresponding required training data and time.

Traditional cryptanalysis regards key extraction as the success of the attack while neural cryptanalysis aims to predict ciphertexts without knowing the key. Traditional cryptanalysis performs delicate manual mathematical analysis  to calculate the influence of the key value to plaintext-ciphertext statistics. Our approach is   automatic and scalable. We further compare them in Table~\ref{tab:comparison}.
\begin{table*}[]
\begin{tabular}{|c|c|c|c|c|c|}
\hline \hline
\textbf{Approaches} & \textbf{Required Data} & \textbf{Required Knowledge} & \textbf{Attack Goal} & \begin{tabular}[c]{@{}c@{}}\textbf{Attack Success}\\ \textbf{Condition}\end{tabular} & \textbf{Attack Enabler} \\ \hline
\hline
\textbf{Cryptanalysis} & \begin{tabular}[c]{@{}c@{}}Plaintext-ciphertext\\ Pairs\end{tabular} & Cipher Algorithm & Key Recovery & \begin{tabular}[c]{@{}c@{}}Attack Complexity\\ $<$ Exhausitive Key\\ Search Complexity\end{tabular} & \begin{tabular}[c]{@{}c@{}}Unbalanced\\ Statistical\\ Property\end{tabular} \\ \hline
\begin{tabular}[c]{@{}c@{}}\textbf{Learning-aided}\\ \textbf{Cryptanalysis (e.g.~\cite{albassal2004neural,martinasek2016profiling})}\end{tabular} & \begin{tabular}[c]{@{}c@{}}Plaintext-ciphertext\\ Pairs\end{tabular} & Cipher Algorithm & Key Recovery & \begin{tabular}[c]{@{}c@{}}Attack Complexity\\ $<$ Exhausitive Key\\ Search Complexity\end{tabular} & \begin{tabular}[c]{@{}c@{}}Unbalanced\\ Statistical\\ Property\end{tabular} \\ \hline
\textbf{Neural Cryptanalysis} & \begin{tabular}[c]{@{}c@{}}Plaintext-ciphertext\\ Pairs\end{tabular} & No Further Knowledge & Ciphertext Prediction & \begin{tabular}[c]{@{}c@{}}Cipher Match Rate\\ $>$ Base Match Rate\end{tabular} & \begin{tabular}[c]{@{}c@{}}Predictability by\\ Neural Network\end{tabular} \\ \hline \hline
\end{tabular}
\caption{Comparison between differential cryptanalysis and neural-cryptanalysis.} 
\label{tab:comparison} 
\end{table*}



The closest related work on neural cryptanalysis is done by Alani~\cite{alani2012neuro, alani2012neuro1}. The author claims to be able to successfully predict plaintexts from ciphertexts of DES and 3-DES by learning from around $2^{11}$ and $2^{12}$ plaintext-ciphertext pairs, respectively. The work adopts a cascade neural network architecture.~\footnote{Alani~\cite{alani2012neuro, alani2012neuro1} reported the average training time of 51 minutes and 72 minutes using MATLAB on a single computer for DES and 3-DES, respectively.} 
However, our experimental findings suggest that the claims in~\cite{alani2012neuro, alani2012neuro1} on full-round DES and 3DES cannot be reproduced (further discussed in Section~\ref{sec:DES3Models}).

Our neural cryptanalysis approach should not be confused with learning-aided cryptanalysis that applies neural networks to improve the statistical profiling in the conventional cryptanalysis (e.g., as in~\cite{albassal2004neural,martinasek2016profiling}). These solutions are still based on traditional cryptanalysis, e.g., requiring the knowledge of cipher algorithms. Our neural cryptanalysis has completely different attack goals and methodology.

Our contributions are summarized as follows.

\begin{itemize}
    \item We present a new methodology to evaluate the strength of cipher primitives with neural networks. Compared with traditional cryptanalysis, it does not rely on the knowledge of cipher algorithms. 
    We define the difficulty of mimicking a cipher by three metrics, cipher match rate, training data, and time complexity, to measure the strength of a cipher.
   
    \item We show the effectiveness of neural cryptanalysis on round-reduced DES and proprietary cipher Hitag2. The experiments show that the strength of Hitag2 is weaker than the 3-round DES in neural cryptanalysis. We discuss the network architecture by applying three different networks.  Experiments show that the most powerful attack neural network varies from cipher to cipher. While a fat-and-shallow shaped fully connected network is the best to attack the round-reduced DES, a deep-and thin shaped fully connected network works best on Hitag2. 
    \item We compare three common activation functions (e.g. sigmoid, relu, tanh) in neural cryptanalysis. Experiments show that Sigmoid converges more quickly than the other two activation functions, which means it can reach certain accuracy with the minimum training time. There is no substantial difference in the converged cipher match rate. 
    We explore the impact of the training data volume on the converged cipher match rate. It shows that more training data significantly improves the converged cipher match rate. With $2^{16}$ Hitag2 training pairs, the neural network achieves a cipher match rate of around $66\%$. When training with $2^{20}$ pairs, the cipher match rate reaches $98\%$.  
    
\end{itemize}

Neural cryptanalysis enables the automatic black-box evaluation of cipher strengths. This evaluation methodology appears quite powerful and exciting, potentially applicable to all ciphers, enabling researchers to compare cipher strengths in a unified framework.



\section{Background and Cipher Mimicking}

 Conventional cryptanalysis heavily relies on unbalanced statistical properties. According to the different ways to find the unbalanced properties, cryptanalysis is categorized into different approaches. Differential cryptanalysis is one of the most common approaches among them. We use it as the representative to introduce conventional cryptanalysis.
 \begin{mydef}[Difference]
  The difference $\Delta$ of two equivalent length bit streams $P_1$ and $P_2$ is defined as $\Delta = P_1 \oplus P_2$ .
\end{mydef}
 
 \begin{mydef}[Differential Path]
  When two plaintexts are encrypted by a multi-round cipher, a $k$-round differential path $\Delta_i \rightarrow \Delta_{i+k}$ is a path where their difference is $\Delta_i$ at the $i$-th round and $\Delta_{i+k}$ at the $(i+k)$-th round, respectively.
 \end{mydef}
 
 \begin{mydef}[Probability of Differential Path $\Delta_i \rightarrow \Delta_{i+k}$]
  Given a fixed path, its probability is the fraction of plaintext pairs that have the difference $\Delta_{i+k}$ at the $(i+k)$-th round among all the pairs having difference $\Delta_i$ at the $i$-th round.
 \end{mydef}
 A conventional differential cryptanalysis approach (e.g.,~\cite{biham1992differential,biham1991differential}) works as follows.

\begin{enumerate}
    \item {\sc Attack path construction phase}: 
     There are two steps: the $p$ rounds and the $q$ rounds. The attacker's goal is to construct a $p$-round differential path with a sufficiently high probability. Given an attacker-chosen plaintext difference $\Delta_{0}$, each possible value of the difference $\Delta_{p}$ at the $p$-th round should occur with the approximately equivalent probability of $2^{-n}$ in a perfect information-theoretically secure cipher, where $n$ is the length of the plaintext-ciphertext. 
    
    
    In reality, when there is a path with a much higher probability than $2^{-n}$, an attacker may leverage it to infer the subkey involved in the last $q$ rounds. $q$ is usually a small value.
    
    \item {\sc Subkey extraction phase}: The attacker is assumed to know many plaintext pairs satisfying the above $\Delta_0$ and their corresponding ciphertext pairs. The high probability of differential path $\Delta_0 \rightarrow \Delta_p$ guarantees that there are much more pairs having the given difference $\Delta_p$ at the $p$-th round than an arbitrary value. All of these ciphertext pairs are partially decrypted to the $p$-th round by a guessed subkey in the last $q$ round. 
    
    \begin{itemize}
        \item   If the guessed subkey is correct, then the calculated differences at $p$-th round would match $\Delta_p$ with the expected high probability.
        \item If the guessed subkey is wrong, then the partial decryption is meaningless. The attacker picks a new guess and repeats.
    \end{itemize} 
   
    \item {\sc Exhaustive search phase}: With the known subkey bits identified in the above phase, the attacker aims to extract all the unknown key bits by trying all the possible key combinations, a partial brute-force approach. 
    The attack complexity required to crack the secret key represents the strength indicator.    
    
\end{enumerate}

Differential cryptanalysis relies on the identified multi-round high-probability differential paths. The entire path space is too large to search exhaustively. Thus, human's intuition and experiences are important. 


\smallskip
\noindent
{\bf Mimicking ciphers by neural networks.}
In neural cryptanalysis, we assess the strength of a cipher by measuring how easy it is for a neural network to mimic the cipher. Intuitively, an attacker should not compute ciphertexts from plaintexts, or vice versa. Thus, a neural network (without the key) should not be able to compute ciphertexts, either. {\em The cipher's strength is reflected by how well an entity (a neural network)  mimics the cipher's operations.}
We test whether or not a neural network trained on plaintext-ciphertext pairs can output the correct ciphertexts on new plaintexts. Intuitively, in our model the easier a cipher algorithm can be mimicked by a well trained neural network, the less secure this cipher is. 

\smallskip
\noindent
{\bf Choice of neural network architectures.}  Different network architectures have different advantages. Long short-term memory (LSTM) is good at predicting sequence dependency relationships while convolutional neural network (CNN) is good at extracting features within a small local area. Neither of them fits the cipher algorithm case, as symmetric ciphers usually apply substitutions and permutations among bits. Fully connectivity is able to represent these relationships between bits. Thus, we choose fully connected neural networks.
    
    Ideally, the relationship between the ciphertext and plaintext should achieve: 1) each bit in the plaintext can have influences on all of the bits in the ciphertext and vice versa; 2) ciphertext is designed to be close to a randomly generated binary stream. Therefore, we decide to choose the multi-layer fully connected neural network as the architecture and the softmax classifier for each bit of the ciphertext stream. The full connection ensures that each bit in the plaintext stream can have an impact on every bit in the ciphertext. 

Besides the regular fully connected neural network, we also evaluate cascade fully connected neural networks mentioned in \cite{alani2012neuro}. The ordinary one only fully connects adjacent layers. The cascade network includes the extra full connection between interval layers. We compare this cascade architecture with the ordinary fully connected architecture.

The ability of neural networks also varies from their scale and shape. Intuitively, a larger neural network with more parameters has a more powerful mimic capability. Is a fat but shallow neural network better or a deep but thin neural network better for the cipher mimic task? To answer this question, we also perform experiments with two neural network models under opposite shape settings. 
    
\smallskip    
\noindent
{\bf Choice of metrics.}  
Given a $n$-bit ciphertext, we count the correct prediction for every single bit independently instead of the entire ciphertext. Therefore, we count the bitwise cipher match rate as the neural network's prediction accuracy.  

The converged training time and required minimum training dataset are also important indicators of the mimic difficulty. According to our measurement, this training complexity varies from different neural network architectures. However, there is no single neural network architecture that always outperforms others across all objectives. Therefore, we decide to quantify the cipher strength as its optimal attack complexity ever achieved. 

\smallskip    
\noindent
{\bf Best-effort strength evaluation}  For each cipher, we apply multiple neural networks in a network suite to mimic it. The best-effort mimicking metrics (i.e. highest accuracy and lowest complexity) are identified as its strength. Similar to traditional case, the strength is a relative notion to the attacking capability -- the neural network suite. We can never conclude a cipher is absolute strong since more powerful attack may appear in the future. Similar limitations also exist for conventional cryptanalysis. The neural network suite indicates the current attacking capability.    

\section{Definitions, Metrics, and Computation }

Symmetric encryption is generally described as the relationship $C=E(P,k)$, where $C$ denotes the ciphertext space, $P$ denotes the plaintext space, $k$ is the secret key, and $E(\cdot)$ is the encryption algorithm. Traditional cryptanalysis evaluates ciphers by attack complexity for cracking the key $k$ with the knowledge of $P$, $C$ and $E(\cdot)$.

In neural cryptanalysis, we assume that $E(\cdot)$ is unknown. Therefore, we transfer the goal as  mimicking the operation $E(\cdot,k)$, instead of cracking key $k$. The difficulty of mimicking algorithm $E(\cdot,k)$ is our security indicator. 
We choose to represent $E(\cdot,k)$ as a trainable neural network. Like an encryption algorithm, this neural network takes plaintexts as input and outputs ciphertexts, without the knowledge of key $k$. Intuitively, stronger ciphers are more difficult for neural networks to mimic.
We formalize neural cryptanalysis as a 4-element tuple $(M_1,M_2,N,S)$ in Definition~\ref{def:neural-cryptanalysis}.  

\begin{mydef}[Neural-cryptanalysis System] A neural cryptanalysis system can be described as a tuple $(M_1,M_2,N,S)$, where
$M_1$ and $M_2$ are two mutual exclusive finite sets of plaintext-ciphertext pairs $(p,c)$ that satisfy $c=E(p,k)$, $p\in Z_2^m$, and $c\in Z_2^n$.  $Z_2$ denotes the binary value space  $\{0,1\}$ and $Z_2^t$ represents a binary stream of length $t$. $N$ is a set of neural networks trained by $M_1$ and tested by $M_2$.
$S$ is the strength indicator generated by training and testing the neural networks in $N$.  
\label{def:neural-cryptanalysis}
\end{mydef}


Our cryptanalysis consists of three operations, {\sc Cipher data collection}, {\sc Mimic model training}, and {\sc Security indicator generation}. We briefly describe the first two operations, as they are straightforward. We present the {\sc Cipher strength quantification} operation in more details.


\smallskip
\noindent
{\sc Cipher data collection} collects required plaintext-ciphertext pairs to train and test the neural networks using the black-box cipher. We collect two sets of plaintext/ciphertext pairs, $M_1$ for training and $M_2$ for testing. 

\begin{figure*}
    \begin{minipage}[b]{.33\linewidth}
    \centering
    \includegraphics[width=\textwidth]{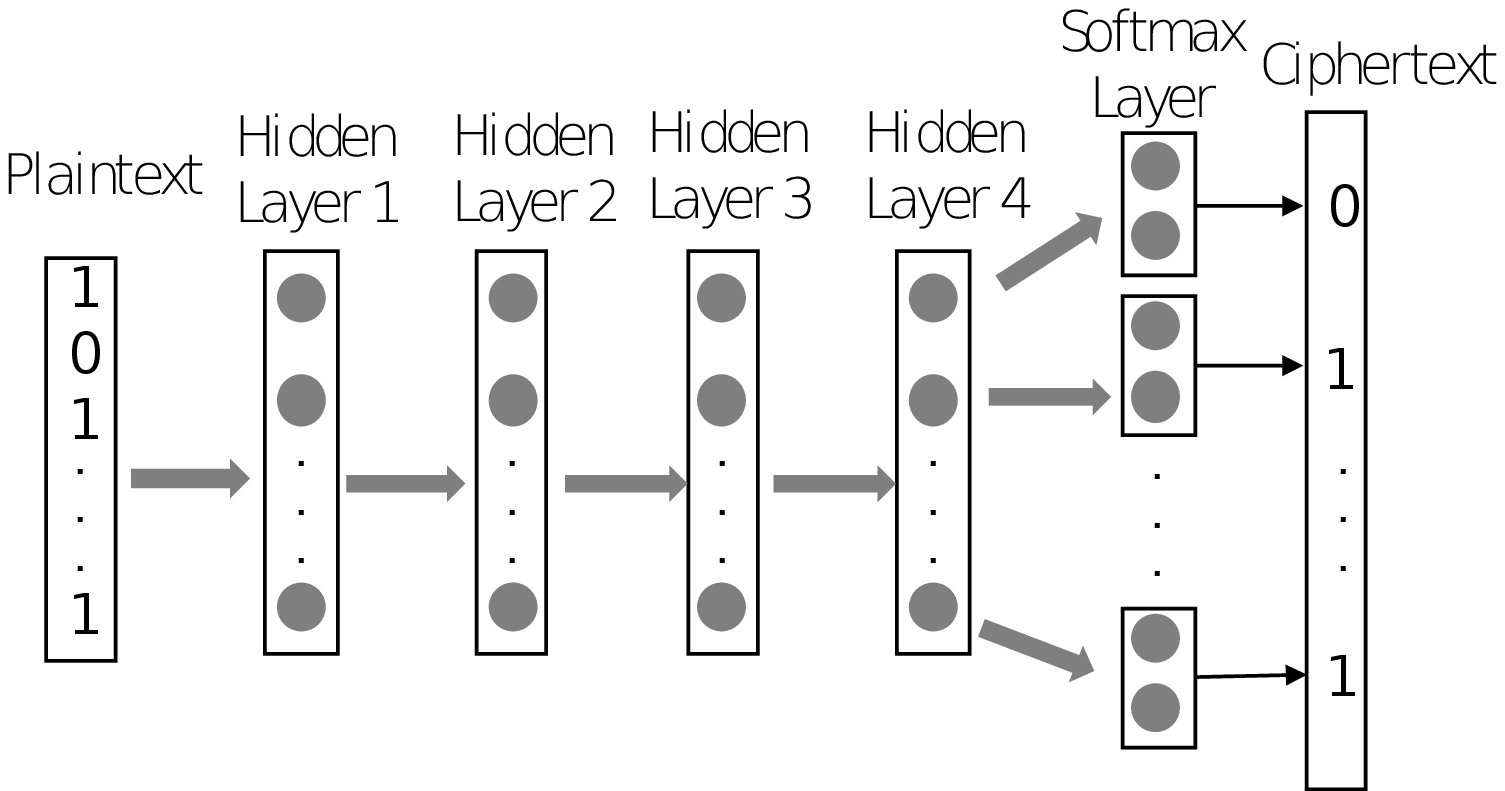}
    \subcaption{Deep and thin network}\label{fig:deep-thin}
    \end{minipage}%
    \begin{minipage}[b]{.33\linewidth}
    \centering
    \includegraphics[width=0.7\textwidth]{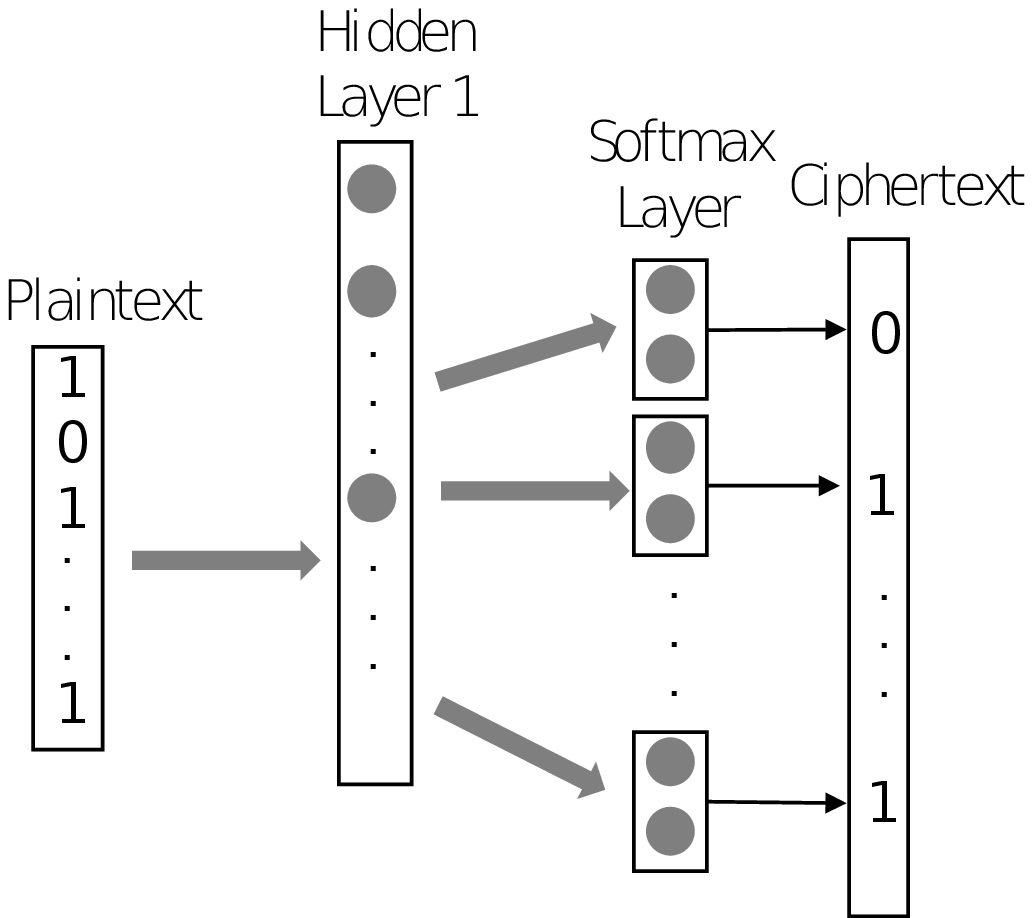}
    \subcaption{Fat and shallow network}\label{fig:fat-shallow}
    \end{minipage}
    \begin{minipage}[b]{.33\linewidth}
    \centering
     \includegraphics[width=\textwidth]{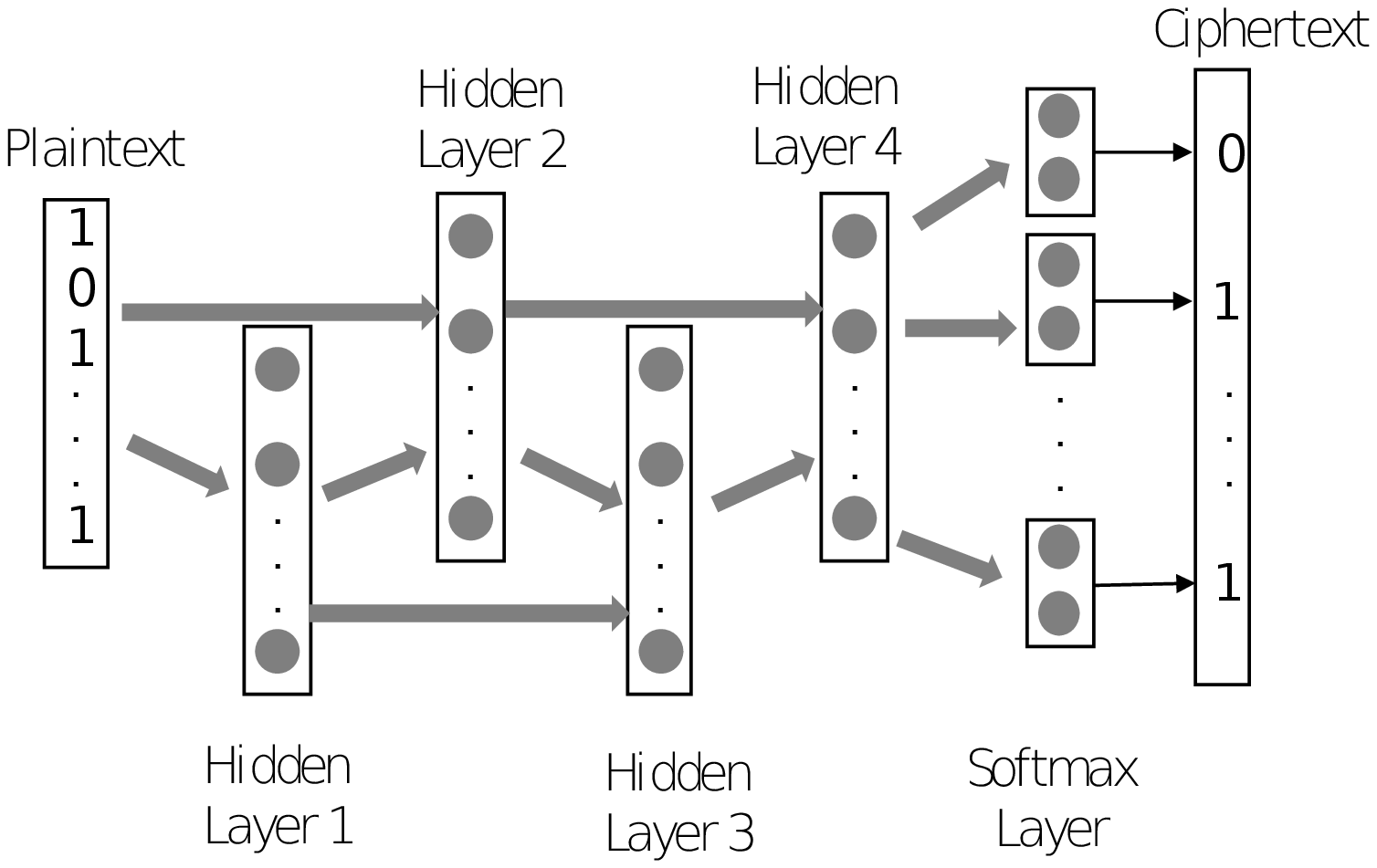}
    \subcaption{Cascade network}\label{fig:cascade}
    \end{minipage}
\caption{Three different neural network architecture applied in experiments }\label{fig:networks}
\label{fig:nn}
\end{figure*}
\smallskip
\noindent

{\sc Mimic model training} trains multiple neural networks to mimic the target cipher and to identify the one with the best performance. 
We train a neural network suite $N$. Each candidate neural network is trained with pairs $M_1$. The neural network suite is the key to the maximum attack capability. 
We include three multi-binary classifier neural networks in the neural network suite in this work.   
The details are shown in Fig.~\ref{fig:nn}. Each cipher bit is predicted by a softmax binary classifier. The entire $n$-bit ciphertext is predicted by $n$ softmax classifier, separately. These three networks differ from the connection and shape. Fig.~\ref{fig:deep-thin} and Fig.~\ref{fig:fat-shallow} are two fully connected neural networks. Fig.~\ref{fig:deep-thin} is a deep and thin network with 4 thin hidden layers. Each layer has 128 neurons. Fig.~\ref{fig:fat-shallow} is a fat and shallow network with 1 fat hidden layer which has 1000 neurons. Fig.~\ref{fig:cascade} is a network with cascade connection. Besides the full connection between every two adjacent layers, every two interval layers are also fully connected. We select one successful size setting from ~\cite{alani2012neuro1}. It has four hidden layers with the size 128, 256, 256, 128 neurons, respectively.


\smallskip
\noindent
{\sc Security indicator generation} computes a security strength metric $S$ composed of mimicking accuracy and complexity. We gradually increase the training data until the mimicking accuracy cannot be further improved. The maximum mimicking accuracy with its minimally required training complexity is used as the final cipher security indicator. If one cipher holds a higher prediction accuracy while the other cipher holds a lower training complexity, our analysis prioritizes the accuracy metric.

The final security indicator $S$ is composed of three elements $(Cmr,Comp_{data},Comp_{time})$, where $Cmr$ is the cipher match rate (Definition~\ref{def:accuracy}), $Comp_{data}$ is the minimum required training data, and $Comp_{time}$ is the converged training time. The last two elements represent the attack complexity. Cipher match rate representing the mimicking accuracy is defined next.


\begin{mydef}[Cipher Match Rate]
 Cipher match rate $Cmr \in [0,1]$ is the bitwise accuracy for the ciphertext prediction of a neural network in neural cryptanalysis. When $m$ $n$-bit ciphertexts are generated by the neural network, $$Cmr = \frac{\sum_{i=0}^{m}\sum_{j\in [0,n-1], c_{ij}=c_{ij}'}1}{m\times n}$$  where $c_{ij}$ and $c_{ij}'$ denote the correct and predicted j-th bit of i-th ciphertext, respectively.
 \label{def:accuracy}
\end{mydef}








We represent the entire evaluation process of neural cryptanalysis in Algorithm~\ref{alg:evaluation}

\RestyleAlgo{ruled}
\begin{algorithm}
\SetAlgoLined
\KwResult{$S=(Cmr,Comp_{data},Comp_{time})$}
 \textbf{Initialization}: Set Hyperparameter $m_1,m_2, Cmr_{base}$
 
 ($2^{m_1}$ is the number of plaintext-ciphertext pairs in $M_1$; $2^{m_2}$ is the number of pairs in $M_2$; $Cmr_{base}$ is the base match rate.)\;
 
 $Cmr = 0, Cmr_{improve}=0, Comp_{data}=m_1-1$\; 
 
 \textbf{Generating $M_2$}: Randomly select $2^{m_2}$ plaintexts and obtain corresponding ciphertexts\;
 
 \While{ ($Cmr \leqslant Cmr_{base}$ or $Cmr_{improve}>0$) and $Comp_{data}<m-m_2$}{
  $Comp_{data} = Comp_{data}+1$\;
  $Cmr_{improve}=0$\;
 \textbf{Generating $M_1$} Randomly select $2^{comp_{data}}$ plaintexts $p\in Z_2^m$ and $p \notin M_2$, obtain corresponding ciphertexts.\;
 \For {each neural network model $n_i \in N$}{ 
  Train $n_i$ to converge with $iter$ iterations\;
  $Comp_{time} = iter$\;
  $Cmr_{n_i}$ = {test bitwise accuracy with $M_2$}\;
  \eIf{$Cmr_{n_i}>Cmr$}{
  $Cmr_{improve} += Cmr_{n_i}-Cmr$;
   $Cmr=Cmr_{n_i}$\;
   }{}
  }
 }
 \caption{Pseudocode for neural cryptanalysis evaluation. Each plaintext is a $m$-bit binary stream.}
 \label{alg:evaluation}
\end{algorithm}


\section{Experimental Evaluation}\label{sec:experiments}

We apply neural cryptanalysis on the round-reduced DES and a real-world widely deployed CPS cipher Hitag2. We also systematically evaluate how various neural network parameters impact the cryptanalysis outcomes and compare the security strengths of Hitag2 and round-reduced DES.

\noindent
{\bf Experimental setup.}
Because the full-round DES algorithm is complex, it is common to first analyze a round-reduced version~\cite{biham1991differential}. Due to its symmetry, there is no difference between predicting ciphertexts from plaintexts and prediting plaintexts from ciphertexts.
 Hitag2 is a stream cipher which generates one secret bit at a time based on a 48 bits state value in a linear-feedback shift register (LFSR). Hitag2 has been reverse engineered in~\cite{hitag2ReverseEngineering}.

Our experiments aim to answer the following questions. 

\begin{enumerate}
    \item How well do the three neural networks mimic cipher algorithms? 
    \item How do the various factors (e.g. network shapes, connections, activation functions, and training data volume) impact the mimicking capabilities?
    \item How strong is the cipher Hitag2 compared with a round-reduced DES?

\end{enumerate}

The three neural networks are implemented with Tensorflow.  We use the sigmoid activation function as the default choice except for the comparative experiments on activation functions. We choose the softmax classifier and the cross-entropy loss function for backward propagation optimization. The maximum training epoch is set to 350. The batch size is 1000. We use a personal laptop to run the training tasks.
We compare the attack capability between the three different neural networks: {\em i)} a fat and shallow neural network, {\em ii)} a thin and deep neural network, and {\em iii)} a cascade neural network from~\cite{alani2012neuro}.

\subsection{Impact of Different Networks on DES}\label{sec:DES3Models}


 To decide whether the attack succeeds, we use a \textit{base match rate} as the baseline accuracy.  In our experiment, when and only when the cipher match rate is higher than the base match rate, the attack is regarded as successful. 
 
\textbf{Base Match Rate of 1-round DES} The base match rate for most ciphers is 50\%, which can be achieved by random guess statistically. However, we set the base match rate as 75\% for 1-round DES. The reason for which is explained next. DES algorithm follows the Feistel structure. In such a structure, cipher bits are separated into two equivalent length parts. In each round, only one part (half block) is updated by the cryptographic round function. The other half is only permuted by bit shifting. Therefore, we assume the half block bits updated by the cryptographic functions is hard to guess while the other half is easy to guess. In that case, the base match rate for the difficult half is still 50\%, however, the base match rate for the simple shifting half is 100\%. Therefore, for the entire 64 ciphertext bits, the base match rate is 75\%.

\begin{figure}
    \centering
    \includegraphics[width=0.5\textwidth]{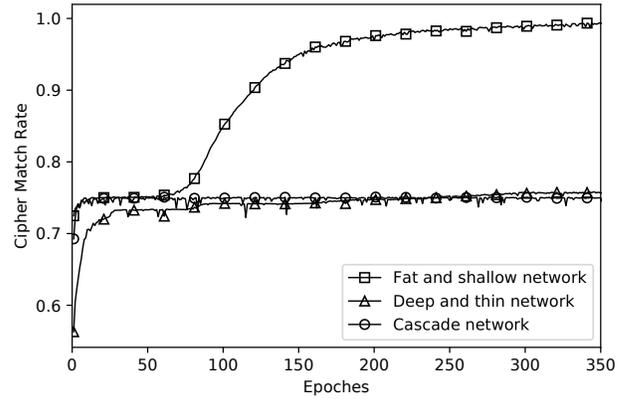}
    \caption{Predicted Accuracy on 1-round DES}
    \label{fig:3models}
    \vspace{-1.5em}
\end{figure}

We implement the round-reduced DES and collect two $2^{17}$ plaintext/ciphertext pairs. Half of them are used to train the three neural networks. The other half is used to test the well-trained model. 
As shown in Fig.~\ref{fig:3models}, the fat and shallow network achieves the highest accuracy. It almost perfectly predicts the ciphertext bits with the cipher match rate of 99.7\%.  The other two neural networks only achieve 75\% cipher match rate, which is no better than the base match rate for one-round DES. It means that the two networks are not better than random guessing for the right 32 bits cryptographic function output. Therefore, we conclude that when the attack object is DES-like algorithms, the fat and shallow shaped network is the best choice among the three neural networks. 

\subsection{Impact of Different Activation Functions on DES}

\begin{figure}
    \centering
    \includegraphics[width=0.5\textwidth]{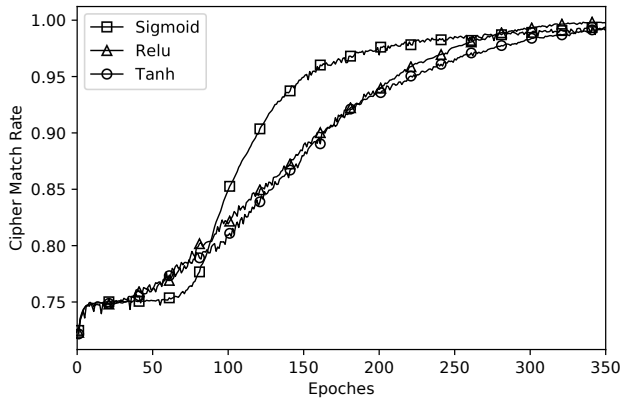}
    \caption{Influence of Activation Functions}
    \label{fig:activation}
    \vspace{-1.5em}
\end{figure}

Activation functions are the only non-linear components of the neural networks. 
There are three common activation functions sigmoid, tanh and relu.  We conduct comparative experiments on the three activation functions. We use the fat and shallow network in the comparative experiments for activation functions.  The result is shown in Fig.~\ref{fig:activation}. They reach similar cipher match rates. However, sigmoid function learns much faster than tanh and relu functions.

\subsection{DES Measurement Results} 
\begin{figure}
    \centering
    \includegraphics[width=0.5\textwidth]{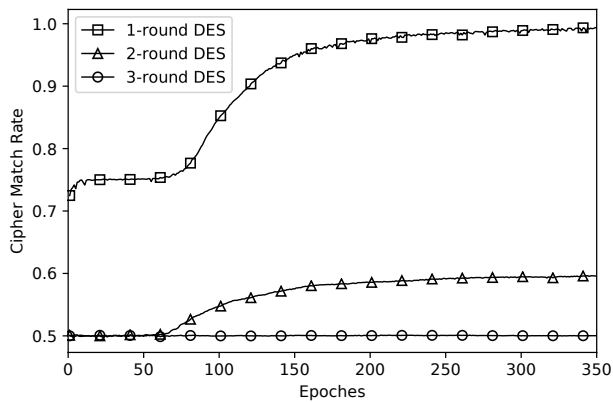}
    \caption{Attack capacity summary for round-reduced DES}
    \label{fig:capacity}
    \vspace{-1.5em}
\end{figure}

Based on the former experiments, we identify the fat and shallow network with sigmoid activation as the optimal choice among these settings for DES. We test higher rounds DES to observe the mimicking capability of this neural network.   
The mimic results for 1-round, 2-round and 3-round DES are shown in Fig.~\ref{fig:capacity}. (Only 1-round DES holds the base match rate as 75\%. 2-round and 3-round DES have the base match rate of 50\%.) The fat and shallow neural network successfully attacked 1-round DES and 2-round DES, with the cipher match rate higher than their base match rate. However, this network cannot mimic 3-round DES successfully. The cipher match rate stays at 50\%. Therefore, this model cannot evaluate ciphers stronger than 3-round DES. The reason is likely the simplicity of the network used in these experiments. To evaluate more complex cipher algorithms, more sophisticated architecture or larger scale neural networks with more parameters are required.  We leave it as future work.  

\subsection{Hitag2 Measurement Results}\label{sec:hitag2model}

Besides the round-reduced DES, we also conduct experiments on Hitag2, which is a widely deployed stream cipher used in modern car key systems. Stream ciphers differ from block ciphers. Block ciphers (e.g. DES) encrypt $n$ bits plaintext into $n$ bits ciphertext with a key. Stream ciphers consistently output an unlimited secret stream bit by bit.  The single output bit is calculated based on an $m$-bit binary stream held in an LFSR.  Then, the state of the LFSR is automatically updated and another secret bit is output based on the updated binary stream in the LFSR. We regard the block cipher algorithm as a $n$ bits to $n$ bits mapping and the stream cipher algorithm as  $m$ bits to 1 bit mapping. Neural cryptanalysis treats the 2 cipher mimicking tasks the same. Hitag2 only outputs 1 bit at a time. Our neural networks only have 1 softmax binary classifier rather than 64 classifiers as in the DES cases. 

We implemented Hitag2 and collected $2^{17}$ input/output pairs. $2^{16}$ pairs are used as training data and $2^{16}$ pairs are used in the evaluation.
We apply the fat and shallow network, deep and thin
network to mimic Hitag2 cipher. The cascade connection architecture is discarded because it does not show any advantage over the ordinary full connectivity architecture in the DES experiments.  
\begin{figure}
    \centering
    \includegraphics[width=0.5\textwidth]{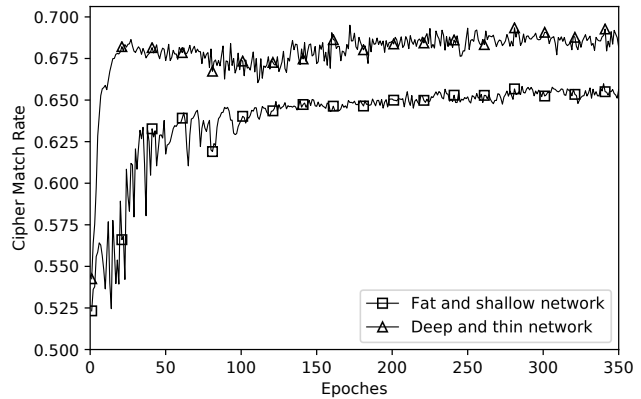}
    \caption{The predicted accuracy of Hitag2 on two models}
    \label{fig:hitag2_model}
    \vspace{-1.5em}
\end{figure}

Figure~\ref{fig:hitag2_model} displays the increasing cipher match rate along the training epochs for Hitag2. 
 With the cipher match rates being higher than 50\%, both the two networks mimic the Hitag2 successfully. The deep and thin network outperforms the fat and shallow one. However, this result is contrary to DES cases in Section ~\ref{sec:DES3Models}. The fat and shallow network shows more power to attack round-reduced DES than the deep and thin network. This observation suggests that there is no single one optimal choice for all ciphers. 

\subsection{Impact of Training Data Size}\label{sec:trainingdata}

We measure whether or not a larger training data set can substantially improve the converged cipher match rate. 
 
\begin{figure}
    \centering
    \includegraphics[width=0.5\textwidth]{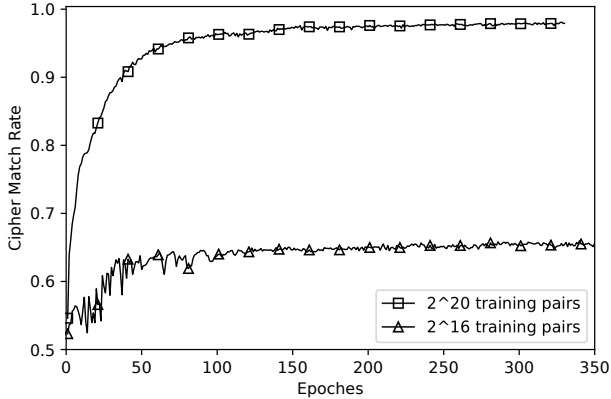}
    \caption{Influence on predicted accuracy by training size}
    \label{fig:hitag2_train_data}
    \vspace{-1.5em}
\end{figure}

 In Section ~\ref{sec:hitag2model}, with $2^{16}$ pairs training data, the maximum cipher match rate of the fat and shallow neural network is around 66\%. 
We increase the available training data to $2^{20}$ pairs and test the converged model on the same test set. Fig.~\ref{fig:hitag2_train_data} shows that when training set increases to $2^{20}$ from $2^{16}$, the cipher match rate rises to around 98\% from 66\%.  

To compare the cipher strengths between two cipher primitives, we maximize their cipher match rates by increasing the training data. The maximum cipher match rate with its data and time complexity are used for the comparison.

\subsection{Security Comparison w.r.t. Round-reduced DES}

\begin{figure}
    \centering
    \includegraphics[width=0.46\textwidth]{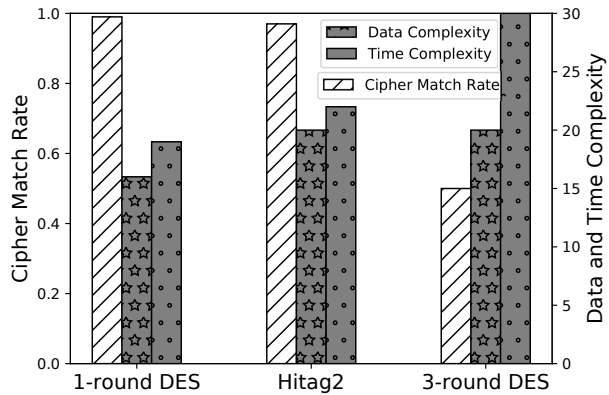}
    \caption{Comparison of Hitag2 and Round-reduced DES. The left y-axis shows the cipher match rate. The right y-axis shows the minimum training data and time complexity required to achieve such a cipher match rate. The data and time complexity is shown in log scale.}
    \label{fig:comparison}
    \vspace{-1.5em}
\end{figure}

We compare the security level of Hitag2 with respect to various round-reduced DES algorithms. As visualized in Fig.~\ref{fig:comparison}, the optimal attack result for 1-round DES is $(99.7\%, 2^{16}, 2^{19})$. The optimal attack result for Hitag2 is $(98\%,2^{20},2^{22})$. The 3-round DES is not attacked successfully with the $2^{20}$ training data and $2^{30}$ iterations.  Therefore the strength of Hitag2 is located between the 1-round DES and 3-round DES, indicating very weak security.



\noindent
{\em Summary of findings.} From these experiments, we observe that the following findings. 
\begin{itemize}
    \item Fully connected neural networks successfully mimic the 1,2-round DES and Hitag2.
    \item For different ciphers, the most powerful mimicking neural networks may be different. This result indicates that one needs to try different neural networks when evaluating a new unknown cipher.
    \item The network with Sigmoid activation function learns faster than Tanh and Relu. However, there is no substantial difference in their converged cipher match rates. 
    \item More training pairs significantly increase the converged cipher match rate as expected.  
\end{itemize}
  
\section{Discussion}

We train neural networks to predict ciphertexts from plaintexts. One can use the approach to predict plaintexts from ciphertexts. It is equivalent to mimic the reverse function of a cipher algorithm. 

{\sc Choice of Base Match Rate} The base match rate is a manually defined baseline of cipher match rate. When the network fails to learn any useful information from the plaintext-ciphertext pairs, the cipher match rate is always around 50\% for a large test set. A higher cipher match rate indicates that useful information to predict the cipher bits is learned from the plaintext-ciphertext pairs.  In the one-round DES case, half cipher bits are updated by the cryptographic function while the other half cipher bits are shifted from the plaintext bits. Thus, we set the base match rate of 1-round DES to be 75\%, instead of 50\%.  

{\sc Limitations and future work} Although successfully mimicking Hitag2, the neural networks we use are simple. These simple neural networks can only evaluate ciphers weaker than 3-round DES. To evaluate more complex or non-deprecated ciphers in the Federal Information Processing Standard (FIPS) Publication 140-2, more powerful neural networks are required. We plan to improve the attack capability in the future work. It involves the following open problems.  


{\sc How to design the neural network to fit cipher mimicking tasks?} Cipher algorithms are composed of iterated arithmetical and logical operations, which are not in the traditional scope of neural networks. Current neural network architectures are not designed for cipher relations.  Fully connected architecture, although capable of mimicking the complicated dependency between ciphertext bits and plaintext bits, might not be the most efficient architecture. Customizing neural networks with fewer parameters to mimic the cipher function is interesting future work. 


{\sc How to train the neural network effectively and efficiently? } Symmetric ciphers are complex functions. To mimic them,  the neural network must include many parameters. However, networks with too many parameters, are difficult to train.  Understanding the impact of loss functions and optimization algorithms is important. Which is the best choice combination for cipher mimic tasks? Can we  customize the loss function and optimization algorithm for this specific task? These questions need to be answered by future work.

\section{Related Work}

The related work can be summarized into two categories, cryptanalysis solutions and neural networks for cryptanalysis research.

\subsection{Cryptanalysis for Symmetric Ciphers}

The specific approaches to evaluate cipher strengths vary dramatically case by case. The representative methods include differential cryptanalysis \cite{biham1992differential}, linear cryptanalysis \cite{matsui1993linear} and their variants. The differential approach attacks DES with $2^{37}$ encryptions and the linear approach attacks DES with $2^{47}$ known-plaintexts complexity, respectively. 


In cyber-physical systems,  the evaluation can only be done with the help of reverse engineering \cite{nohl2008reverse}. Many real-world CPS ciphers have been shown broken. The widespread DST40 cipher which is often used in immobilizer transponders was reverse-engineered and found broken \cite{bono2005security}.  Another widely used CPS cipher, Hitag2, was also cracked after reverse-engineering \cite{verdult2012gone,tekian2014doctoral,garcia2016lock1}. CPS cipher Megamos Crypto was found to be vulnerable after reverse-engineering and cryptanalysis \cite{verdult2015dismantling,verdult2015cryptanalysis}. The analysis found that there is a flawed key generation which can reduce the exhaustive search complexity from $2^{96}$ to $2^{57}$.  Those flaws in CPS ciphers have affected many real-world systems \cite{garcia2016lock1}.

\subsection{Neural Networks in Cryptanalysis}

 The applications of deep learning in cryptanalysis are limited. One major direction is to the learning-aided side channel attacks. \cite{hospodar2011machine,lerman2014power,martinasek2016profiling,cagli2017convolutional}. Researchers apply learning approaches to replace the traditional statistical profiling phase in side-channel attacks. This approach differs greatly from our approach. The deep learning approach only acts as one component of the entire side-channel attack.  Therefore, the analysis still requires the cipher algorithm. 
 
 There are also a few applications in traditional cryptanalysis which do not rely on side-channel information.  
 \cite{chandra2007applications} leverages neural networks to classify ciphertexts generated from different cipher algorithms. They successfully distinguish the ciphertexts of Enhanced RC6 and SEAL.
\cite{albassal2004neural} uses neural networks to determine the correct key in the key recovery attack of traditional cryptanalysis. Their attack is successful on the 2-round and 3-round HypCipher. However, similar to the learning-aided side-channel attacks, it does not change the white-box algorithm assumption which the traditional attack relies on. 

The work most relevant to ours is  \cite{alani2012neuro,alani2012neuro1}.  The work treats DES as a black-box.  A cascade neural network structure is applied to directly take the plaintexts as input and trained to output corresponding ciphertexts. The author claimed to have successfully attacked DES and 3-DES. However, the results are not reproducible by us. In addition, the parameters of their networks are too few to approximate the full DES and 3-DES. Our work presents a universal metric to enable the direct comparison between ciphers, which is new. Applying neural cryptanalysis to CPS ciphers is also new. 

\section{Conclusions}

In this paper, we proposed a new approach to evaluate the strength of symmetric ciphers  in a black-box fashion without knowing the specific algorithm. We defined quantitative metrics to capture a cipher's strength. These metrics are calculated from the difficulty to attack the cipher via neural networks.  

Using Hitag2 and round-reduced DES algorithms, we experimentally demonstrated that our metric and methodology are practical and useful to quantify cipher strengths and allow one to directly compare between ciphers.
We showed how various factors associated with the neural networks can impact the analysis outcomes. 
This new neural cryptanalysis approach has the potential to automate the security evaluation for cipher systems. 
\vspace{-1em}
\bibliographystyle{IEEEtran}
\bibliography{Ref}

\end{document}